\documentclass[fleqn,usenatbib]{mnras}

\usepackage{newtxtext,newtxmath}

\usepackage{float}

\usepackage[T1]{fontenc}
\newcommand{\unit}[1]{{\ \rm #1}}
\DeclareRobustCommand{\VAN}[3]{#2}
\let\VANthebibliography\thebibliography
\def\thebibliography{\DeclareRobustCommand{\VAN}[3]{##3}\VANthebibliography}

\usepackage{graphicx}	
\usepackage{amsmath}	
\usepackage{subfigure}
\usepackage{multirow}
\usepackage{multicol}
\usepackage{arydshln}
\usepackage{cite}
\usepackage{threeparttable}
\usepackage{bm}
\usepackage{tikz,xcolor,hyperref}

\definecolor{lime}{HTML}{A6CE39}
\DeclareRobustCommand{\orcidicon}{
	\begin{tikzpicture}
	\draw[lime, fill=lime] (0,0) 
	circle [radius=0.16] 
	node[white] {{\fontfamily{qag}\selectfont \tiny ID}};
	\draw[white, fill=white] (-0.0625,0.095) 
	circle [radius=0.007];
	\end{tikzpicture}
	\hspace{-2mm}
}
\foreach \x in {A, ..., Z}{%
	\expandafter\xdef\csname orcid\x\endcsname{\noexpand\href{https://orcid.org/\csname orcidauthor\x\endcsname}{\noexpand\orcidicon}}
}

\title[Counting Method to Constrain Kilonova Anisotropy]{On using the counting method to constrain the anisotropy of kilonova radiation
}

\author[Zhang et al.]{
Siqi Zhang$^{1,2}$\orcidA{},
Furen Deng$^{1,2}$\orcidB{},
Youjun Lu$^{1,2}$\orcidC{}
\thanks{E-mail: luyj@nao.cas.cn}
\\
$^{1}$National Astronomical Observatories, Chinese Academy of Sciences, 20A Datun Road, Beijing 100101, China;\\
$^{2}$School of Astronomy and Space Science, University of Chinese Academy of Sciences,
19A Yuquan Road, Beijing 100049, China
}

\date{Accepted XXX. Received YYY; in original form ZZZ}

\pubyear{2024}

\begin{document}

\label{firstpage}
\pagerange{\pageref{firstpage}--\pageref{lastpage}}
\maketitle

\begin{abstract}
A large number of binary neutron star (BNS) mergers are expected to be detected by gravitational wave (GW) detectors and the electromagnetic (EM) counterparts (e.g., kilonovae) of a fraction of these mergers may be detected in multi-bands by large area survey telescopes. For a given number of BNS mergers detected by their GW signals, the expected numbers of their EM counterparts that can be detected by a survey with given selection criteria depend on the kilonova properties, including the anisotropy. In this paper, we investigate whether the anisotropy of kilonova radiation and the kilonova model can be constrained statistically by the counting method, i.e., using the numbers of BNS mergers detected via GW and multi-band EM signals. Adopting simple models for the BNS mergers, afterglows, and a simple two (blue and red)-component model for kilonovae, we generate mock samples for GW detected BNS mergers, their associated kilonovae and afterglows detected in multi-bands. By assuming some criteria for searching the EM counterparts, we simulate the observations of these EM counterparts and obtain the EM observed samples in different bands. With the numbers of BNS mergers detected by GW detectors and EM survey telescopes in different bands, we show that the anisotropy of kilonova radiation
and the kilonova model can be well constrained by using the Bayesian analysis. Our results suggest that the anisotropy of kilonova radiation may be demographically and globally constrained by simply using the detection numbers of BNS mergers by GW detectors and EM survey telescopes in multi-bands.
\end{abstract}




\section{Introduction}
\label{sec:intro}
Mergers of binary neutron stars (BNSs) or neutron star-black hole binaries (NSBHs) are predicted to eject neutron-rich material and produce ``kilonovae'' phenomena \citep[][]{LiandPaczyski1998,Metzger2010,Kasen2017} by the heating from the radiative decay of heavy elements formed via the rapid neutron capture process (\emph{r}-process) in the ejecta \citep[e.g.,][]{Lattimer1974}. This prediction has confirmed by the joint detection of the gravitational wave (GW) signal from the merger of a BNS (GW170817) by the advanced Laser Interferometer Gravitational wave Observatory (LIGO) and Virgo \citep{Abbott2017a} and its associated electromagnetic (EM) counterparts, i.e., the short gamma-ray burst (sGRB), its afterglow, and kilonova \citep{Goldstein2017, Savchenko2017, Hallinan2017, Alexander2017, Arcavi2017, Pian2017}. The discovery of GW170817 marks the beginning of the multi-messenger astronomy by using both the GW and EM signals to study the physical processes in strong gravity field and extreme physical environments \citep{Abbott2017b}.

LIGO and Virgo have detected more than $90$ merger events of compact binaries \citep[][]{ O3b2021}, among them only two are BNS mergers, i.e., GW170817 and GW190425 \citep{Abbott2017a, GW190425}. It is expected that many more BNS mergers will be detected by future GW detectors, including the LIGO Voyager \citep{Voyager2020}, Einstein Telescope \citep[ET;][]{ET2010} and Cosmic Explorer \citep[CE;][]{CE2019}. In the coming fourth and fifth observation runs (O4 and O5) of LIGO/Virgo/KAGRA, it is anticipated that $\sim 10^1-10^3$ of BNS mergers will be detected per year \citep{Petrov2022,Weizmann2023}. In the era of the third generation GW detectors, it is expected that upto $\sim10^3$ and $\sim10^4-10^5$ of BNS mergers per year will be detected by LIGO Voyager and CE/ET, respectively \citep[e.g.,][]{Belgacem2019, 2022Singh, mahao2022}. 

The EM counterparts of some BNS mergers detected by GW detectors are also expected to be detected by the time domain large sky surveys, such as the Rubin Observatory Legacy Survey of Space and Time \citep[hereafter Rubin;][]{LSST2009,LSST2019}, the Zwicky Transient Facility \citep[ZTF;][]{Bellm2019, Graham2019}, the Panoramic Survey Telescope and Rapid Response System \citep[Pan-STARRS;][]{Chambers2016}, and the SiTian Projects \citep[SiTian;][]{SiTian2021}. Several works have investigated the detection rates of kilonovae and the afterglows of short Gamma Ray Bursts (sGRBs) associated with BNS mergers, and found that about $20$, $2$, $0.1$, or $7$ ($800$/$50$/$1$/$2000$) kilonovae (afterglows) can be discovered by Rubin, ZTF, Pan-STARRS, or SiTian \citep{Zhu2021b, KNrate1, KNrate2, KNrate3, Yujiming2021, Roberts2011}. Currently only GW170817 has both the GW and the EM (afterglow/kilonova) observations. There are indeed some sGRBs that have afterglow observations with kilonova signatures \citep{GRB050709, GRB060614, GRB070809, GRB130603b_1, GRB130603B_2, GRB150101B, GRB160821B_1, GRB160821B_2, GRB160821B_3}, of which the GW signals could not be detected because no GW detector worked at the merger time of these sGRBs. One may thus expect to detect a large number of kilonovae and sGRBs associated with BNS merger GW events in the near future.  Some of the detected kilonovae may have detailed multi-epoch multi-band observations as GW170817, which enables detailed studies of each one. Some of them may not have detailed observations because of their low luminosity or faint magnitude, or the long cadence time (defined as the interval between consecutive observations of the same sky area chosen from different detection strategy) \citep{KNrate3, Andreoni2022}. The samples of BNS mergers, kilonovae, and sGRBs as a whole may also offer demographic studies on the intrinsic properties of these systems and relevant underlying physics, as well as the individual ones with detailed observations. 

Kilonova radiation is anisotropic because of different properties of the ejecta at different directions. The neutron rich material ejected at directions around the equator has larger opacity and thus emits relatively redder radiation (red component), while that at directions close to the polar has smaller opacity and emits relatively bluer radiation (blue component) \citep[e.g.,][]{Metzger2019, Villar2017}. This difference is the main reason for the anisotropy of the kilonova radiation. Many efforts have been made to study such kilonova anisotropy in the literature. \citet{Metzger2014} first performed an axisymmetric hydrodynamic simulation to generate the light curves of two-component kilonova. Such a two-component model is required for successfully fitting the light curves of AT2017gfo \citep{Chornock2017, Cowperthwaite2017, Coulter2017, Kasen2017, Villar2017}. Furthermore, the light curves of AT2017gfo may be better fitted by a three-component model with a purple component adding into the two-component model \citep{Villar2017, Perego2017}. Recently, many useful simulations and tools were proposed and developed to study the anisotropic for kilonovae, such as the Los Alamos National Laboratory (LANL) grid of radiative transfer kilonova simulations and the time-dependent multi-dimensional Monte Carlo radiative transfer code POlarization Spectral Synthesis In Supernovae (POSSIS) \citep[e.g.,][]{LANL1,LANL2,Bulla2019,Dietrich2017a,Bulla2023,Heinzel2021}. According to these studies of kilonova anisotropy, the observational properties of a kilonova observed at different viewing angle are expected to be different \citep[e.g.,][]{Darbha2020, ZCY2023}. Therefore, it is possible to study the anisotropy of kilonova radiation by using the statistical distribution of kilonova properties obtained from observations if a large number of kilonovae could be detected. 

A bayesian analysis to infer the width of sGRB jets from BNS mergers was proposed by \citet{Farah2020} and the method is denoted as the ``counting method''. Based on the number of sGRB and GW detections, they showed that the posterior distribution of the effective angular width for the sGRB jets can be well constrained by assuming a binomial distribution of the likelihood. Recently, \citet{Chen2023} also adopted a similar method to constrain the geometry of fast radio bursts (FRB). The reason why sGRB and FRB jet can be estimated by this method is that the emission of them is highly anisotropic. Compared with the sGRB and FRB, the emission from a kilonova is relatively less anisotropic, however, the anisotropy may be still significant. It is possible to adopt the ``counting method'' to obtain constraint on the anisotropy of kilonova emission from a large sample of kilonovae observations. The counting method requires little observational information, and it is expected to use for a quick analysis of the large amount of data obtained by the gradually upgraded GW and optical detectors in the future. 

In this paper, we investigate whether the counting method can give some constraints on the structure parameters of kilonova according to simple kilonova and afterglow models of BNS mergers. The paper is organized as follows. In Section~\ref{sec:model}, we introduce the structured two-component kilonova model and the afterglow model. In Section~\ref{sec:method} and Section~\ref{sec:results}, we describe the counting method and perform simulations to demonstrate that the counting method can put substantial constraints on the aniostropy of the kilonova radiation. Conclusions and Discussions are given in Section~\ref{sec:con}.

\section{Kilonova and sGRB models}
\label{sec:model}

In this section, we introduce simple models for kilonovae and sGRB afterglows, respectively. With these models, we can estimate the light curves for any given kilonova and sGRB.

\subsection{Kilonova}
\label{sec:kn} 

\begin{figure}
\centering
\includegraphics[width=0.3\textwidth]{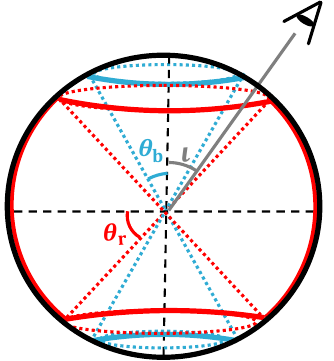}
\caption{
Schematic diagram of the two-component model for kilonovae. The blue component ejecta is around the polar direction with an opening angle of $\theta_{\rm b}$, and the red component ejecta is around the merger plane with an opening angle of $\theta_{\rm r}$. The viewing angle $\iota$ is defined as the angle between the line of sight (LOS) and the normal of the BNS orbit. 
}
\label{fig:str}
\end{figure}

We adopt a simple two-component model for kilonovae, of which the diagram is schematically shown in Figure~\ref{fig:str}. In this model, a kilonovae is axi-symmetric and can be characterized by two components, i.e., a blue component around the polar angle with opening angle of $\theta_{\rm b}$ and a red component around the equator with opening angle of $\theta_{\rm r}$. Note that three-component model or even more complicated models which are calculated by simulations such as POSSIS and LANL were proposed in the literature, but the fittings to the kilonova light curves by using different models only show small differences \citep[see][]{Villar2017, Perego2017, Darbha2020, LANL1,LANL2,Bulla2019}. For the purpose of demonstration and simplification in this paper we adopt the two-component semi-analytic model. We calculate the anisotropic radiation from any kilonova according to the two-component model as those in the previous studies \citep{Darbha2020, Villar2017, Kasen2017, Metzger2010, Korobkin2012}.

A small amount of neutron material can be ejected out from the merger of a BNS, in which the \emph{r}-process nucleosynthesis leads to the formation  of radioactive heavy elements. These heavy elements are unstable and quickly decay to release a large amount of energy, leading to the kilonova phenomenon \citep{Metzger2010}. The temporal evolution of the radioactive heating rate for each component can be approximated as \citep{Korobkin2012}
\begin{equation}
L^{\rm{eff}}_{\rm{in}}(t)=4\times 10^{18}m^{\rm{eff}}_{\rm{ej}}\times \left[0.5-\pi^{-1}\arctan\left(\frac{t-t_0}{\sigma}\right)\right]^{1.3}\rm{erg\,s^{-1}},
\label{eq1}
\end{equation}
where $t_0=1.3$\,s, and $\sigma=0.11$, $m^{\rm{eff}}_{\rm{ej}}$ is the effective mass of the \emph{r}-process nucleosynthesis ejecta, which equals the ejecta mass of the blue or red component divided by a solid angle factor of $(1-\cos\theta_{\rm b})$ or $\sin\theta_{\rm r}$. 

A fraction of the total radioactive decay power $L^{\rm{eff}}_{\rm{in}}$ in the ejecta may lead to the thermalization of the ejected material and be transferred into the kilonova radiation \citep{Metzger2010}, i.e.,
\begin{equation}
\epsilon_{\rm{th}}(\emph{t})=0.36\left[\rm{exp}(-\emph{a} \emph{t})+\frac{\ln\left(1+2\emph{b} \emph{t}^{\emph{d}}\right)}{2\emph{b} \emph{t}^{\emph{d}}}\right],\label{eq2}
\end{equation}
where $a$, $b$, and $d$ are fitting constants. These constants are functions of the mass and velocity of the ejecta as seen in \citet[][see their Table 1]{Barnes2016}. 

After considering the effective radioactive power $L^{\rm{eff}}_{\rm{in}}$ and the thermalized fraction $\epsilon_{\rm{th}}$, we take the form of the common analytic solution derived by \citet{Arnett1982}, and then obtain the output effective bolometric luminosity as
\begin{equation}
L^{\rm{eff}}_{\rm{bol}}(t)={\rm exp}\left(\frac{-t^2}{t^2_{\rm{d}}}\right)\times \int_0^t L^{\rm{eff}}_{\rm{in}}(t')\epsilon_{\rm{th}}(t'){\rm exp}\left(\frac{t'^2}{t^2_{\rm{d}}}\right)\frac{t'}{t_{\rm{d}}}d\frac{t'}{t_d}.
\label{eq3}
\end{equation}
Here $t_{\rm{d}}=(2\kappa m^{\rm{eff}}_{\rm{ej}}/\beta v_{\rm{ej}} c)^{1/2}$, with $\beta =13.4$ denoting a shape constant, $v_{\rm{ej}}$ representing the mass and velocity of r-process for red or blue ejecta, $c$ is the speed of light, and $\kappa$ denoting the opacity for the red or blue component.

Due to the large opacity of the ejecta, the radiation from kilonovae can be approximated as blackbody radiation. In order to simulate the multi-band light curve, each component is assumed to have a black-body photosphere with constant radius expansion \citep{Arnett1982}. At any given time, the temperature of each component is defined by its bolometric luminosity and radius, using the Stefan–Boltzmann law. After the ejecta cool to a critical temperature ($T_{\rm c}$) at which the photosphere begins to recede into the ejecta, then the temperature remains constant. The temperature and radius of the photosphere for different component are given by
\begin{equation}
T_{\rm phot}(t)={\rm max}\left[\left(\frac{L^{\rm{eff}}_{\rm{bol}}(t)}{4\pi\sigma_{\rm{SB}}v_{\rm ej}^2t^2}\right)^{1/4},T_{\rm c}\right]\label{eq4},
\end{equation}
and
\begin{eqnarray}
R_{\rm{phot}}(t)=
\begin{cases}
v_{\rm{ej}}t,      & \textrm{if}\,\,\left(\frac{L^{\rm{eff}}_{\rm{bol}}(t)}{4\pi \sigma_{\rm{SB}} v_{\rm{ej}}^2t^2}\right)^{1/4}>T_{\rm{c}}, \\
\left(\frac{L^{\rm{eff}}_{\rm{bol}}(t)}{4\pi \sigma_{\rm{SB}}T_{\rm{c}}^4}\right)^{1/2},   & \textrm{if}\,\,\left(\frac{L^{\rm{eff}}_{\rm{bol}}(t)}{4\pi \sigma_{\rm{SB}} v_{\rm{ej}}^2t^2}\right)^{1/4} \leq T_{\rm{c}}. \label{eq5}
\end{cases}
\end{eqnarray}
It is worth to mention that the $T_{\rm{phot}}$ and $R_{\rm{phot}}$ do not change with the $L^{\rm{eff}}_{\rm{bol}}$ scaled to the radiation inside the cone. Therefore, we can easily get the radiation from the core for each component by these parameters.

\subsubsection{Anisotropic flux}

The actual area of radiation received by the detector is equivalent to the projection of the radiation region of red and blue component on the plane which normal to the LOS. To calculate the anisotropic flux, we divide the angle of the radiation part into small solid angles as $d\Omega$.
We define unit vector $\bm{a}$ as the normal to each $d\Omega$ and $\bm{q}$ as the unit vector pointing to the observer.
Since we can only receive the part of emission which facing to us, the condition $\bm{q}\cdot \bm{a} > 0$ is required. 
The anisotropic radiation for each component can be written as
\begin{equation}
\label{eq6}
F_{\rm{cp}}(\nu,t) =   B_{\rm{cp}}(\nu,t)\left(\frac{R_{\rm{phot}}}{d_{\rm L}}\right)^2\iint_{(\bm{q}\cdot \bm{a})>0}(\bm{q}\cdot d\bm{\Omega}),
\end{equation}
where $\nu$ is the frequency, $B(\nu)$ is the intensity of blackbody emission which is calculated from the $T_{\rm{phot}}$ described in  Equation (\ref{eq4}), $d_{\rm L}$ is the luminosity distance and the subscript ``$\rm {cp}$'' denotes different component of the kilonova. We obtain the total flux $F({\nu},t)=\sum_{\rm cp} F_{\rm cp} (\nu, t)$ by summing those from different components and convert it to the AB magnitude as
\begin{equation}
\label{eq7}
m_{\rm{AB}}(t)=-2.5\log_{10}\left(\frac{(z\!+\!1)\!\!\int \!F(\frac{\nu}{1+z},t) R(\nu)d\ln\nu}{\int \!R(\nu)d\ln\nu}\right)-48.6 ,
\end{equation}
where $R(\nu)$ is the transmission for different filters and $z$ is the redshift.

\begin{figure*}
\centering
\includegraphics[width=0.9\textwidth]{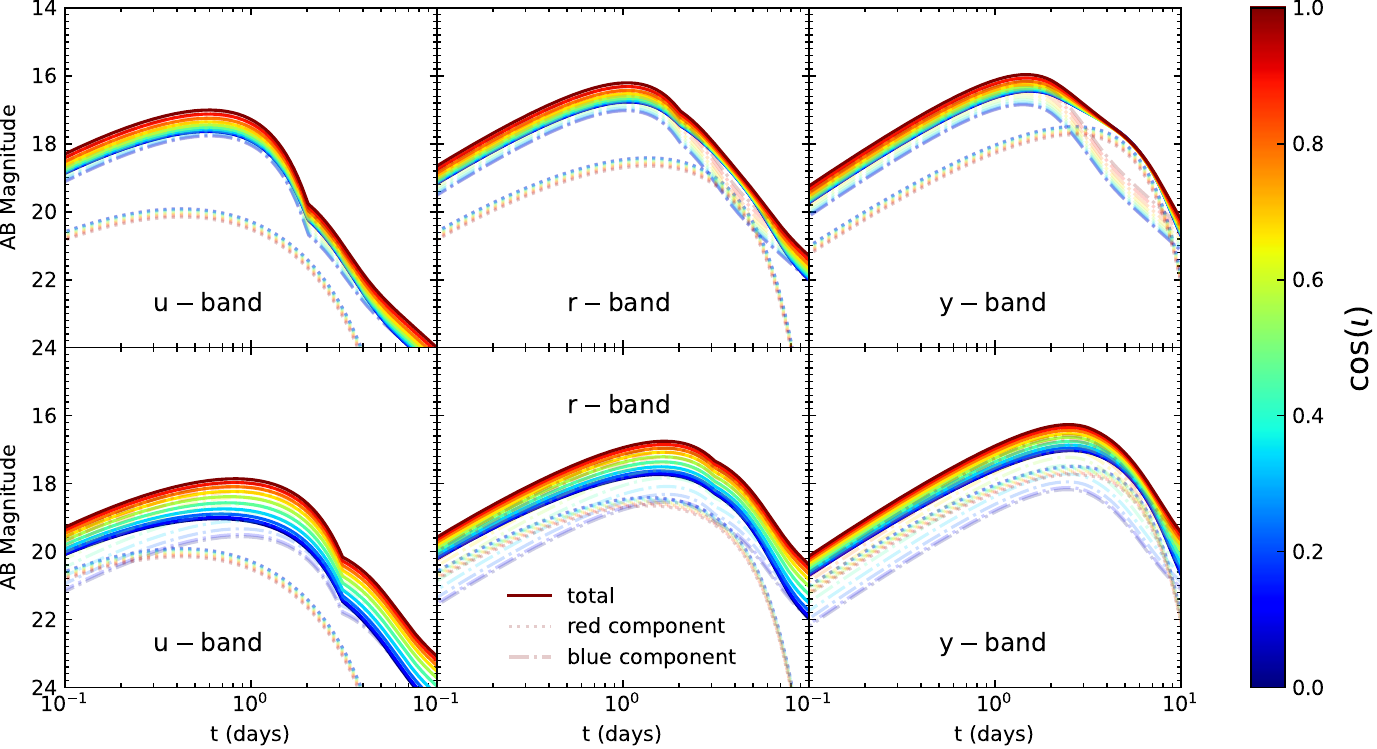}
\caption{
Light curves of kilonovae viewing at different viewing angles in different bands. Top and bottom panels show the cases with input parameters of $(m_{\rm{ej}},\theta_{\rm b},\theta_{\rm r},z)=(0.05 M_\odot,\pi/3,\pi/3,0.005)$ and $(0.05 M_\odot,\pi/6,\pi/3,0.005)$, respectively. Solid, dashed, and dotted dash lines represent the light curves, and the contributions to them from the red component and the blue component, respectively.  Colors of the curves represent the viewing angles as indicated by the colorbar at the right side.
}
\label{fig:knlc} 
\end{figure*}

Figure~\ref{fig:knlc} shows some example light curves for kilonovae in the u-, r-, and y-bands viewing at different $\iota$. As seen from this Figure, the light curves increases at the beginning, reach their peaks around $0.5$-$2$\,days after the BNS merger, and then decrease with elapsing time $t$. After the peak, there is a bump caused by the red component, evolving slower than the blue component. Since the blue component distributes around the poles and the red component distributes near the equator, the blue component dominates the kilonova flux received by an observer viewing it at small $\iota$ and the bump after the peak, due to the red component, is less obvious; the red component is prominent when $\iota$  is large, and the late bump is more obvious. The light curve in each band depends on $\iota$ because of the anisotropic nature of kilonovae, and this dependence in the u-band is more stronger than that in the y-band. If $\theta_{\rm b} = 60^\circ$ or $30^\circ$ but $\theta_{\rm r}$ fixed at $60^\circ$, the peak magnitude of the light curve in the u-/y-band is expected to change by upto $0.78$/$0.55$ or $1.33$/$0.90$ magnitudes with $\iota$ changing from $0^\circ$ to $90^\circ$.

\subsection{Afterglow of GRB}
\label{sec:ag} 

GRBs can have afterglows in a wide range of bands due to the interaction of their jets with the surrounding interstellar medium (ISM), which creates internal and external counter-shocks and thus accelerates electrons to produce synchrotron radiation \citep{Meszaros1997}. Afterglow emission depends on the structure of GRB jets and the environment of GRBs \citep[e.g.,][]{Rossi2002, Zhang2002}. \citet{Lazzati2018} found that a Gaussian structure jet model can well match the observations of the GRB (GRB170817A) and afterglows associated with GW170817. Here we adopt the Gaussian structure jet model, for which the viewing-angle-dependent energy can be described as
\begin{equation}
\label{eq8}
E(\vartheta)=E_0\exp\left(-\frac{\vartheta^2}{2\vartheta_{\rm c}^2}  \right),
\end{equation}
where $E_0$ is the on-axis equivalent isotropic energy, $\vartheta_{\rm c}$ represents the typical opening angle of the jets.

\begin{figure}
\centering
\includegraphics[width=0.45\textwidth]{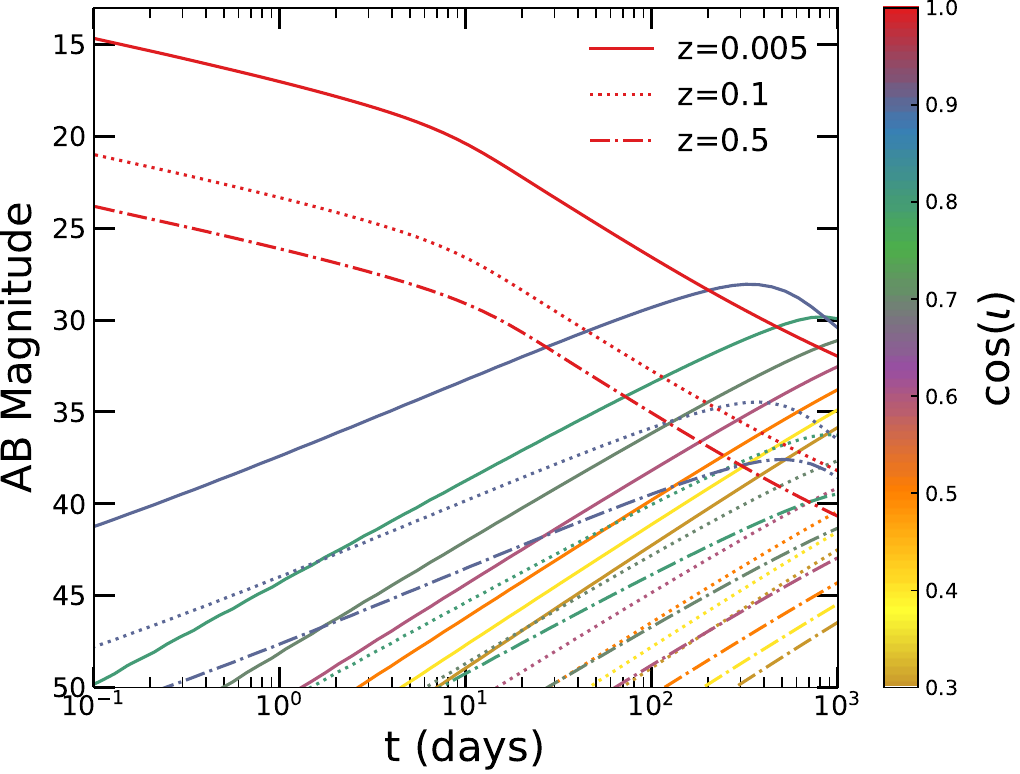}
\caption{
The u-band light curves from the afterglow produced by a Gaussian structured jet viewing at different angles. Solid, dashed, and dotted-dashed lines show cases with redshift $z=0.05$, $0.1$, and $0.5$, respectively. The color of the curves represents the viewing angle of the afterglow as indicated by the right colorbar. The parameters are similar with the afterglow of GRB170817A.  
}
\label{fig:aglc}
\end{figure}

We adopt the python package \emph{afterglowpy} \citep{afterglowpy} to calculate the light curves of afterglows. We adopt the model parameters for afterglows given by the best fit to GRB170817A observations \citep{Troja2018}, as listed in Table~\ref{par of ag}. Figure~\ref{fig:aglc} shows example light curves in the u band obtained by observers viewing the afterglows of a sGRB with redshift $0.005$ at different viewing angles. As seen from this Figure, the peaks of the light curves decrease sharply with increasing $\iota$ especially at small viewing angles. The light curves of afterglows in other bands have similar evolution pattern and only u band is shown in Figure~\ref{fig:aglc}. 

\begin{table}
\centering
\caption{%
The afterglow model parameters for GRB170817A
}

\setlength{\tabcolsep}{0.4mm}{		
\begin{tabular}{cccccccc}		\hline \hline 
 $\log E_0$	&$\vartheta_{\rm c}$	&$\vartheta_w$	&$\log n_0$	& $p$	&$\log \epsilon_e$	&$\log \epsilon_B$	 \\
	\noalign{\smallskip}\hline\noalign{\smallskip}		
 52.73 & 0.057 & 0.62 & -3.8 & 2.155 & -1.51 & -3.20  \\
\noalign{\smallskip}\hline \hline
\end{tabular}
}
\begin{tablenotes}
\footnotesize
\item Notes: $E_0$ is the on-axis equivalent isotropic energy, $\vartheta_{\rm c}$ is the opening angle of the jets, $\vartheta_w$ is a truncation angle, $n_0$ is the number density of the interstellar medium, $p$ is the power-law index of accelerated shock, $\epsilon_e$ is the magnetic field energy fraction and $\epsilon_B$ is the accelerated electron energy fraction. We adopt the best fit values for these model parameters given by \citet{Troja2018}.
\end{tablenotes}

\label{par of ag}
%
\end{table}

\section{Method}
\label{sec:method}

In this section, we introduce the counting method and make some simulations to limit the parameters of kilonova. Simulation results for four different scenarios are also obtained in this section.

We adopt a Bayesian method to statistically constrain the parameters of kilonovae in this paper, which is similar to the one by \citet{Farah2020} for inferring the width of sGRB jets from BNS mergers. The model of the kilonova is more complicated compared with that for the sGRB afterglow. The expected anisotropy of kilonova emission could be considerable, but is not as significant as that for sGRBs. The kilonova light curves at different bands depend on the viewing angle differently, thus multi-band observations encode important information of the anisotropy. Therefore, it is possible to constrain the anisotropy of kilonova radiation by using multi-band observations. For some individual kilonovae with detailed light curve observations, one may fit the multi-band light curves by using sophisticated kilonova model to reveal the anisotropic nature of kilonova radiation, as that done for GW170817 \citep[e.g.,][]{Perego2017, Villar2017, Drout2017, Zhu2021, ZCY2023}, with which one could obtain strong constraints on related physical parameters.  For those kilonovae without detailed or even no light curve observations, however, one may still use the information of multi-band ``observed'' and ``unobserved'' kilonovae to constrain their properties via statistical methods, such as the counting method. 

Several parameters that are mostly relevant to the counting method are described as follows.
\begin{itemize}
\item The total number of BNS merger events being detected by GWs and EM waves (either afterglows or kilonovae) in any given band with given S/N thresholds are denoted as $N_{\rm{GW}}$ and $N_{\rm{EM}}$, respectively. The number of EM counterparts detected in different bands can be different, and we use the vector $\bm{N_{\rm{EM}}}$ to denote the different numbers of detected EM counterparts in different bands. In this paper, for illustration purpose, we adopt the six bands of Rubin for our calculation, i.e., u, g, r, i, z, and y bands, therefore $\bm{N_{\rm{EM}}}$ is written as  $\bm{N_{\rm{EM}}}=\{N_{\rm{EM,u}}, N_{\rm{EM,g}}, N_{\rm{EM,r}}, N_{\rm{EM,i}}, N_{\rm{EM,z}}, N_{\rm{EM,y}}\}$. 
\item The viewing angle $\iota$ is defined as the angle between the LOS and the normal of the BNS orbit.
\item The total mass of kilonova ejecta is denoted as $m_{\rm{ej}}$. We set the mass ratio (denoted as $\chi$) of the blue component to the red component at a fixed value of $0.46$ \citep[][]{Villar2017} in our calculations. In principle, this parameter can have a wide distribution among different kilonovae, of which the effect will be discussed in Section~\ref{sec:con}.
\item The fraction of the GW events that have detected EM counterparts  is denoted as $q_x=N^x_{\rm{EM}}/N_{\rm{GW}}$ for one band in $\{u,g,r,i,z,y\}$.  
\end{itemize}

The goal of this paper is to investigate whether the anisotropy of kilonava radiation can be constrained by using the counting method, i.e., constraining $\theta_{\rm b}$, $\theta_{\rm r}$, and $m_{\rm{ej}}$ via $N_{\rm{GW}}$ and $\bm{N_{\rm{EM}}}$ based on some mock observations. According to the Bayesian theorem, one can obtain the posterior probability distribution 
\begin{multline}
P(m_{\rm{ej}},\theta_{\rm b},\theta_{\rm r}|N_{\rm{GW}},\bm{N_{\rm{EM}}})\\
=\prod \limits_{x}^{n=6}\frac{P(N_{\rm{GW}}, N_{\rm{EM}}^x|m_{\rm{ej}},\theta_{\rm b},\theta_{\rm r})P(m_{\rm{ej}},\theta_{\rm b},\theta_{\rm r})}{P(N_{\rm{GW}},N_{\rm{EM}}^x)},
\label{eq9}
\end{multline}
where the superscript `$x$' represents any given band, and $n$ indicate the total number of bands being considered, and it equals to six for Rubin in our calculation. We assume that the detection in each band is independent, ignoring that the observations in different bands may not be independent from each other with a certain observation strategy. With this assumption, the information from multi-bands may be cumprod in the likelihood. In the model, there is naturally a correlation among detection numbers of the events observed in different bands due to the fact that, we use different bands to observe the same mock sources which are calculated by a given set of model parameters. Therefore, the correlation between the detections in different bands is already included in the calculation of $P(N_{\rm{GW}}, N_{\rm{EM}}^x|m_{\rm{ej}},\theta_{\rm b},\theta_{\rm r})$. However, one should note that the searches of the EM counterparts in different bands may be not independent and the correlations among different bands may be determined by the searching strategy. In this paper, we do not consider this in the likelihood for simplicity. We also assume uniform priors for parameters $m_{\rm{ej}}$, $\theta_{\rm b}$, and $\theta_{\rm r}$, therefore, the posterior is proportional to the likelihood. Since $q$ can be calculated from these three parameters, the likelihood for one band can be rewritten as $P(N_{\rm{GW}},N_{\rm{EM}}^x|q_x(m_{\rm{ej}},\theta_{\rm b},\theta_{\rm r}))$. We assume that the GW sources are independent from each other, then the likelihood should be a binomial distribution which can be written as
\begin{multline}
P(m_{\rm{ej}},\theta_{\rm b},\theta_{\rm r}|N_{\rm{GW}},\bm{N_{\rm{EM}}})=\\
\prod \limits_{x}^{n=6}\frac{q_x(m_{\rm{ej}},\theta_{\rm b},\theta_{\rm r})^{N_{\rm{EM}}^x}{(1-q_x(m_{\rm{ej}},\theta_{\rm b},\theta_{\rm r}))^{N_{\rm{GW}}-N_{\rm{EM}}^x}}}{\iiint q_x(m_{\rm{ej}},\theta_{\rm b},\theta_{\rm r})^{N_{\rm{EM}}^x}{(1-q_x(m_{\rm{ej}},\theta_{\rm b},\theta_{\rm r}))^{N_{\rm{GW}}-N_{\rm{EM}}^x}} dm_{\rm{ej}}d\theta_{\rm b}d\theta_{\rm r}}.\\
\label{eq10}
\end{multline} 
Therefore, how to calculate $q_x(m_{\rm{ej}},\theta_{\rm b},\theta_{\rm r})$ in each band is the key to obtaining the likelihood function. The detection rate of EM counterparts in one band is given by
\begin{equation}
q_x=P(A_{{\rm{EM}},x}=1|A_{\rm{GW}}=1), 
\label{eq11}
\end{equation}
where $A$ is a bool indicator, with $A_{{\rm{EM}},x}=1$ and $A_{\rm{GW}}=1$ for the detection of EM and GW, respectively, $A_{\rm{EM}}=0$ for the non-detection of EM. The detection rate is also related to redshift $z$ and inclination angle $\iota$. For GW, it has additional parameters $\Psi=\{\theta,\phi,\psi\}$. $\theta$ and $\phi$ denote the source's ecliptic colatitude and longitude, $\psi$ is the polarization angle defined by \citet{Apostolatos1994} denoting the orientation of the source relative to the detector. Equation~\eqref{eq11} can be then marginalized as
\begin{equation}
q_x=\iiint P(A_{{\rm{EM}},x}=1|z,\iota)P(z,\iota,\Psi|A_{\rm{GW}}=1)dzd\iota d\Psi\label{eq12}.
\end{equation}
We define a kilonova as observable, i.e., $P(A_{\rm{EM,i}}=1|z,\iota)$, if its peak magnitude is brighter than the threshold of the detection magnitude
\begin{equation}
P(A_{{\rm{EM}},x}=1|z,\iota)=\Theta(m_{{\rm{thr}},x}-m_x(\iota,z|m_{\rm{ej}},\theta_{\rm b},\theta_{\rm r})),
\label{eq13}
\end{equation}
where $\Theta$ is a step function, $m_x$ is the peak magnitude of a kilonova at a given band, which is related to the total mass of the ejecta $m_{\rm{ej}}$, the opening angles of the blue and red components $\theta_{\rm b}$, $\theta_{\rm r}$. $m_{{\rm{thr}},x}$ is the threshold for the $5 \sigma$ limiting magnitude of Rubin in different band. Table~\ref{Rubin} lists the limiting magnitudes for the six Rubin bands. There are also other surveys, such as the Nancy Grace Roman Space Telescope \citep[RST;][]{WFIRST}, Euclid \citep[Euclid;][]{Euclid} and several telescopes summarized above that could be used to find kilonovae with different strategies in the future, however for demonstrate, we only use Rubin in our calculation.

\begin{table}
\centering
\caption{The $5\sigma$ limiting magnitude ($m_{\rm lim}$) of Rubin for six different bands. }
\begin{tabular}{lllllll} \hline \hline
Filter & u & g & r & i & z & y \\ \hline
$m_{\rm lim}$ & 23.87 & 24.82 & 24.36 & 23.93 & 23.36 & 22.47 \\ \hline \hline
\end{tabular}
\label{Rubin}
\end{table}

The term $P(\iota,z,\Psi|A_{\rm{GW}})$ in Equation~(\ref{eq12}) can be written as
\begin{multline}
P(\iota,z,\Psi|A_{\rm{GW}})=\frac{P(\iota,z,\Psi,A_{\rm{GW}})}{P(A_{\rm{GW}})}\\
=\frac{P(\iota,z,\Psi)P(A_{\rm{GW}}|z,\iota,\Psi)}{\iiint P(A_{\rm{GW}}|z,\iota,\Psi)P(\iota,z,\Psi)dzd\iota d\Psi}\label{eq14},
\end{multline}
where $P(A_{\rm{GW}}|z,\iota,\Psi)=\Theta(\rho(z,\iota,\Psi)-\rho_{\rm{\rm{thr}}})$, with the signal to noise ratio (SNR) of the GW signal given by \citep{Finn1993}
\begin{equation}
\rho(\Psi, z,\iota)=8\Theta_{\rm{ap}}\frac{r_{\rm{0}}}{d_{\rm{L}}(z)}\left(\frac{\mathcal{M}_z}{1.2M_\odot}\right)^{\frac{5}{6}}\sqrt{\mathcal{Z}(f_{\rm{max}})}\label{eq15}.
\end{equation}
Here 
\begin{equation}
\Theta_{\rm{ap}} =2[F_+^2(1+\cos^2\iota)^2+4F_\times^2\cos^2\iota]^\frac{1}{2}
\label{eq16},
\end{equation}
with the antenna pattern functions
\begin{eqnarray}
F_+=\frac{1}{2}(1+\cos^2\theta)\cos2\phi\cos2\psi-\cos\theta \sin2\phi \sin2\psi, \\
F_\times=\frac{1}{2}(1+\cos^2\theta)\cos2\phi \sin2\psi+\cos\theta \sin2\phi \cos2\psi,
\label{eq17and18}
\end{eqnarray}
These angles are independent with each other, therefore we can randomly generate points for $\cos{\theta}, \phi/\pi$, $\psi/\pi$, $z$ and $\cos{\iota}$ to decided the SNR for each GW.

In Equation (\ref{eq15}), $r_{\rm{0}}$ represents the characteristic distance parameter for a GW detector and we set $r_{\rm{0}}=1527\unit{Mpc}$ for ET, $\mathcal{M}_z$ is the observed  chirp mass (i.e., $\mathcal{M}_z=\mathcal{M}_{\rm{0}}(1+z)$), $\mathcal{Z}(f_{\rm{max}})$ is a dimensionless function, reflecting the overlap between the GW signal and the effective bandwidth of a GW detector, which increases monotonically as a function of the maximum orbital frequency $f_{\rm{max}}$. \citet{Taylor2012} estimated that $\mathcal{Z}(f_{\rm{max}})$ is close to unity. Therefore, we set $\mathcal{Z}(f_{\rm{max}})=1$ for our following calculations. When LIGO and Virgo is used for calculation, a large number of GWs cannot be detected within the redshift of 0.5. This result leads to a very high value of $q$, which affects the estimation of parameters using our method. Meanwhile, the strong detection ability of ET make the value of $q$ become very small, because many GW can be detected at high redshift, while the corresponding optical signals are basically not detected for now. To balance this two aspects, we choose ET as the GW detector and raise the GW detection threshold to $\rho_{\rm{thr}}=21$ in our calculation.

Substituting Equations~\eqref{eq13} and \eqref{eq14} into Equation~\eqref{eq12}, we have 
\begin{equation}
\begin{aligned}
&q_x(m_{\rm{ej}},\theta_{\rm b},\theta_{\rm r})=\\
&\frac{\iiint \Theta(m_{\rm{thr,x}}-m_x(\iota,z|m_{\rm{ej}},\theta_{\rm b},\theta_{\rm r}))\Theta(\rho(\iota,z,\Psi)-\rho_{\rm{thr}})p(\iota,z,\Psi)dzd\iota d\Psi}{\iiint \Theta(\rho(\iota,z,\Psi)-\rho_{\rm{thr}})p(\iota,z,\Psi)dzd\iota d\Psi }.
\label{eq19}
\end{aligned}
\end{equation}
According to Equation~\eqref{eq19}, we generate $M=10^5$ mock events with different $z$, $\iota$, $\theta$, $\phi$, $\psi$ via Monte-Carlo (MC) method. The values of $\cos{\iota}$, $\cos{\theta}$, $\phi/\pi$ and $\psi/\pi$ are assumed to be randomly distributed in the range of $[-1,1]$. We set the maximum value for redshift as $0.5$ for convenience, as kilonovae at lower redshifts are easier to be detected. For the distribution of $z$, we consider that the merger rate of BNS evolves with redshifts. The number density per unit time for BNS mergers is given by \citep{Chen2022, Zhao2021}
\begin{equation}
p(z)=\frac{d\dot{N}_{\rm{BNS}}}{dz}=\frac{R(z)}{1+z}\frac{dV_{\rm{c}}}{dz},
\label{eq20}
\end{equation}
where $R(z)$ is the BNS merger rate density and $V_{\rm{c}}$ is the comoving volume.  We adopt the results of BNS merger rate density by the model ``$\alpha10$.kb$\beta0.9$'' in \citet{Chu2022}.  By adopting a Lambda Cold Dark Matter ($\Lambda \rm{CDM}$) universe with $H_{\rm{0}}=70\unit{km s^{-1} Mpc^{-1}}$, and $\Omega_{\rm{m}}=0.3$, $\Omega_{\rm{\Lambda}}=0.7$, one can obtain the comoving volume element 
\begin{equation}
\label{eq21}
\frac{dV_{\rm{c}}}{dz}=\frac{c}{H_{\rm{0}}}\frac{4\pi D^2_{\rm{L}}}{(1+z)^2\sqrt{\Omega_\Lambda+\Omega_{m}(1+z)^3}}.
\end{equation}
The MC approximate shows as
\begin{multline}
q_x(m_{\rm{ej}},\theta_{\rm b},\theta_{\rm r})=\\
\frac{\sum_{j=1}^M \Theta(m_{\rm{thr,x}}-m_x(\iota_j,z_j|m_{\rm{ej}},\theta_{\rm b},\theta_{\rm r}))\Theta(\rho(\iota_j,z_j,\Psi_j)-\rho_{\rm{thr}})}{\sum_{j=1}^M \Theta(\rho(\iota_j,z_j,\Psi_j)-\rho_{\rm{thr}}) }.
\label{eqq3}
\end{multline}
The masses of the ejecta for different kilonovae may be different, which depend on the total mass and mass ratio of each BNS merger. The distribution of such masses can be given by numerical simulations \citep[e.g.,][]{Radice2018, Dietrich2017a, Dietrich2017b, Hotokezaka2013} or from the fittings to the observations of kilonovae (such as GW170817, see \citealt{Villar2017}). For simplicity, in the present paper, we assume a simple distribution for the mass of the kilonova ejecta, i.e., either an uniform distribution in the range from $\log m_1$ to $\log m_2$ or a log-normal distribution with a mean and a standard deviation of $\mu$ and $\sigma$, respectively. Besides the distributions of the kilonova mass ejecta, $\theta_{\rm b}$ and $\theta_{\rm r}$, there are some other kilonova parameters (e.g., the ejecta mass ratio $\chi$, and $v_{\rm{ej}}$, $\kappa$, $T_{\rm c}$ for each component). For demonstration purposes, we fix the values of these parameters as those obtained by \citet{Villar2017} in our calculations and give a further discussion on the influence that may be caused by the distribution of these parameters in Section~\ref{sec:con}. 

Given the distribution of the kilonova mass ejecta, $\theta_{\rm b}$, $\theta_{\rm r}$, and other kilonova model parameters, we can generate mock samples and obtain the $q_x$ value [see Eq.~\eqref{eqq3} in different bands, rewritten as $q_x(m_1, m_2,\theta_{\rm b},\theta_{\rm r})$ or $q_x(\mu,\sigma,\theta_{\rm b}, \theta_{\rm r})$]. To generate mock observations, we firstly simulate a number of BNS mergers, and calculate the expected S/N of the GW signal from each BNS merger. For any input parameter set $I = \{m_1, m_2, \theta_{\rm b}, \theta_{\rm r}\}$ or $\{\mu, \sigma, \theta_{\rm b}, \theta_{\rm r}\}$, we can also generate the kilonova signal for each source. Combined with the detection capabilities of EM and GW detectors, we can  obtain the expected $N_{\rm{GW}}$ and $\bm{N_{\rm{EM}}}$ according to the simulations. After having the dependence of $q_x$ with four parameters and the mock observation numbers, we are expected to use Equation~\eqref{eq10} to re-constrain the input parameters. One may keep in mind that the expected numbers of detectable events depend on the searching strategy of the EM counterparts. Therefore, we consider four different cases as listed below.
\begin{itemize}
\item[1)] ``Rubin'': we consider the third generation GW detector ET for the GW detection and the Rubin with the 5$\sigma$ limiting magnitudes given by its wide-field sky survey for searching kilonovae. The EM counterpart model considers kilonova-only.
\item[2)] ``Rubin+'': same as ``Rubin'' listed in the above item but with the $5\sigma$ limiting magnitudes one magnitude deeper than those for ``Rubin'' to demonstrate the effect of detector sensitivity on the constraint.
\item[3)] ``KN+AG\&info'': same as ``Rubin+'' but considering the influence of observations of identified afterglows in the counting. 
\item[4)] ``KN+AG\&noinfo'': same as ``KN+AG\&info'' but without information of whether the EM counterparts are kilonovae or afterglows.

For all the above scenarios, we further consider the influence of the observation strategy. A kilonova may be not detectable even if the peak magnitude of the light curve is above the threshold because the detection capability is influenced by different telescope detection strategies. Following \citet{Zhu2021b}, for demonstration purpose, we simply adopt a factor to describe this effect for the detection probability in our calculations as 
\begin{equation}
\label{eq23}
\min \left[\frac{\Omega_{\mathrm{cov}}}{\Omega_{\mathrm{sph}}}, \frac{\Omega_{\mathrm{FoV}} \dot{t}_{\mathrm{ope}} \min \left(t_{\mathrm{cad}}, \Delta t_i\right)}{\Omega_{\mathrm{sph}}\left(n_{\exp } t_{\exp }+t_{\mathrm{oth}}\right)}\right]
\end{equation}
where $\Omega_{\rm{sph}}=41252.96 \unit{deg^2}$ is the total sky area, $\Omega_{\rm{cov}}$, $\Omega_{\rm{FoV}}$, $\dot{t}_{\rm{ope}}$, $t_{\rm{cad}}$, $\Delta t_i$, $n_{\rm{exp}}$, $t_{\rm{exp}}$, $t_{\rm{oth}}$ are the sky coverage, field of view (FoV), average operation time per day, the optimal interval between consecutive observations of the same sky area, the length of time when a kilonova is above the limiting magnitude of telescope, the condition that the detection of the kilonova by serendipitous search is exposure on at least two filters, the optimal exposure time, time spent for each visit, respectively, for a transient survey. For Rubin, we have $\Omega_{\rm{cov}}=20000 \unit{deg^2}$, $\Omega_{\rm{FoV}}=9.6 \unit{deg^2}$, $\dot{t}_{\rm{ope}}=6 \unit{hr\, day^{-1}}$, $t_{\rm{cad}}=1 \unit{day}$, $\Delta t_i$, $n_{\rm{exp}}=2$, $t_{\rm{exp}}=30 \unit{s}$, $t_{\rm{oth}}=15 \unit{s}$. For a small part of well-localized sources, Rubin will likely be used to focus on their areas to get more details of their EM signals via the Target of Opportunity (ToO) observation strategy \citep{Andreoni2022, Margutti2018}. In this paper, we try to demonstrate a general method which only needs the information about whether the EM counterparts are detected, for which the Wide Fast Deep (WFD) survey strategy can be adopted \citep{Chase2022, KNrate3, Sagus2021, Setzer2019, Andreoni2022b, Scolnic2018}. We do not intend to use the detailed information that may be obtained for those EM counterparts obtained by the ToO strategy for a small number of well-localized sources. For more complex and detailed strategies on searching kilonovae, for example, see \citet{Andreoni2022}, \citet{Almualla2021},  \citet{Frostig2021}, and \citet{Sagus2021}. 
\end{itemize}
Moreover, to illustrate the robustness of our results, we use different distribution of $m_{\rm{ej}}$ in mock observations and reconstruction. In mock observations we still use the log-uniform distribution which is controlled by two parameters $m_1$ and $m_2$ (i.e., with the input parameter set $I = \{m_1, m_2, \theta_{\rm b}, \theta_{\rm r}\}$) to generate samples while in our reconstruction we assume the $m_{\rm{ej}}$ obeys log-normal distribution which controlled by $\mu$ and $\sigma$ (i.e., consider $q_x(\mu, \sigma, \theta_{\rm b}, \theta_{\rm r})$ in posterior probability distribution).

\section{Results}
\label{sec:results}

In this section, we present the main results on constraining the anisotropy of kilonova emission obtained by using number counting, i.e., the number of GW events $N_{\rm{GW}}$ and the number of GW-associated EM counterparts (kilonovae or afterglows) ($\bm{N_{\rm EM}}$) observed in the same sky area(s).

\subsection{$q_x$ and the mock observations}
\label{sec:qx and N} 

\begin{figure*}
\centering\includegraphics[width=1.0\textwidth]{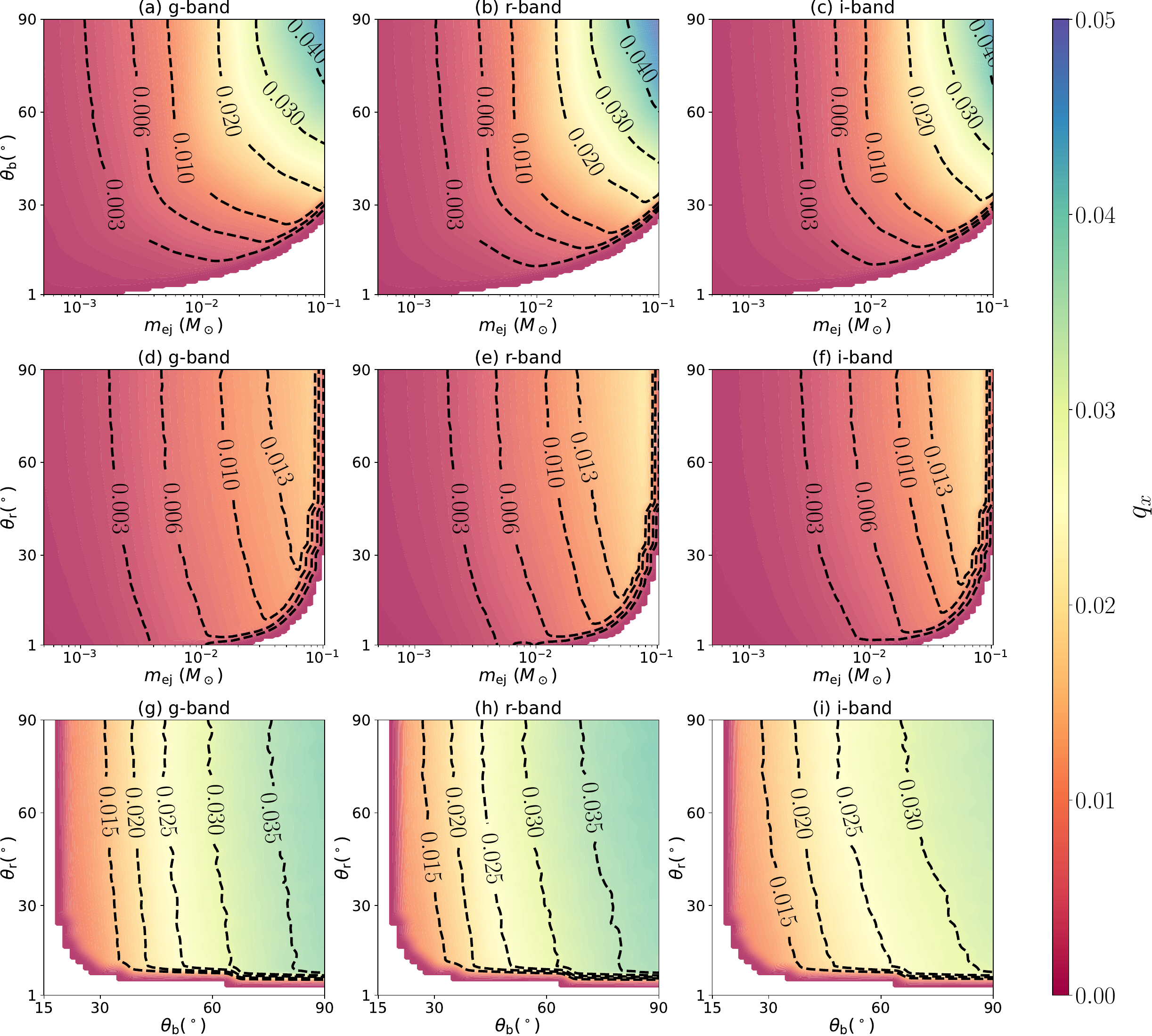}
\caption{
Distributions of $q_x(m_{\rm{ej}},\theta_{\rm b},\theta_{\rm r})$ in the $\theta_{\rm b}$ versus $m_{\rm ej}$ plane (top panels), $\theta_{\rm r}$ versus $m_{\rm ej}$ plane (middle panels), and $\theta_{\rm b}$ versus $\theta_{\rm r}$ plane (bottom panels) for three most sensitive bands of EM detector  g-, r-, and i-bands (from left to right). The values of $\theta_{\rm r}$ for top panels are fixed at $\pi/3$, the values of $\theta_{\rm b}$ for middle panels are fixed at $\pi/6$, and the values of $m_{\rm{ej}}$ are fixed at $0.05 M_\odot$ for bottom panels. Different colors in this figure represent different values of $q_x$ as indicated by the right colorbar, which is obtained from mock observations for BNS merger GW events with specific settings on the corresponding kilonova parameters under the ``Rubin+'' scenario. 
}
\label{fig:q}
\end{figure*}

We investigate the dependence of $q_x$ on four parameters [i.e., $q_x(m_1, m_2,\theta_{\rm b},\theta_{\rm r})$ and $q_x(\mu,\sigma,\theta_{\rm b},\theta_{\rm r})$] for four different cases. For simplicity, we consider the cases with fixed $m_{\rm ej}$ (not a distribution of $m_{\rm ej}$) to illustrate the dependence of $q_x$ on the ejecta mass. Figure~\ref{fig:q} shows $q_x(m_{\rm{ej}},\theta_{\rm b},\theta_{\rm r})$ obtained from our calculations on the plane of $m_{\rm ej}-\theta_{\rm b}$ (top panels), $m_{\rm ej} - \theta_{\rm r}$ (middle panels), and $\theta_{\rm b} - \theta_{\rm r}$ (bottom panels), respectively, for the case ``Rubin+''. In the top, middle, or bottom panels, we fix $\theta_{\rm r}=\pi/3$, $\theta_{\rm b}=\pi/6$, or $m_{\rm{ej}}=0.05 M_{\odot}$. We consider all the u, g, r, i, z, and y bands of ``Rubin+'' observations, and show the results in three most sensitive bands for Rubin, i.e., g, i, and y bands in the left, middle, and right panels, respectively, for brevity. As seen from the top or middle panels in this figure, when $m_{\rm{ej}}$ is higher than $1\times10^{-2}M_{\odot}$, the changes of $\theta_{\rm b}$ or $\theta_{\rm r}$ can bring to the change of $q_x$ about $0.02$ or $0.004$. In the bottom panels, it is also clear that different values of $\theta_{\rm b}$ lead to significant different results on $q_x$. The dependence of $q_x$ on different parameters and its different behaviours at different bands justify that the counting method can be used to constrain kilonova parameters and the anisotropy of kilonova radiation and the combination of multi-band observations would lead to improvement to the constraints on the model parameters comparing with those by single band observations. However, one should note that the induced change of $q_x$ by the change of $\theta_{\rm r}$ is significantly small compared with those by the change of $\theta_{\rm b}$ or $m_{\rm{ej}}$. The reason is that the peaks of kilonova light curves are mainly contributed by the radiation from their blue components, which were used to find them and count the detection numbers. Therefore, the constraints on $\theta_{\rm r}$ by the counting method should be less tight comparing with those on $\theta_{\rm b}$, as will be seen in the next Section for various cases.

We generate mock samples of BNS mergers and kilonovae from it. For demonstration purpose, here we set the input values as $I=\{m_1=10^{-4} \unit{M_{\odot}}, m_2=10^{-1} \unit{M_{\odot}},\theta_{\rm b}={\pi/6},\theta_{\rm r}={\pi/3}\}$. According to the mock samples, we can obtain $N_{\rm{GW}}$ and $\bm{N_{\rm EM}}$. We summarize the estimated numbers in Table~\ref{mock} for four different cases.

\subsection{Reconstruction and constraining capability on the anisotropy parameters}
\label{sec:reconstruction}

We can adopt the Markov Chain Monte Carlo (MCMC) method to constrain the anisotropy of the kilonova model, mainly described by $\theta_{\rm b}$ and $\theta_{\rm r}$, as well as other model parameters, once we have the dependence of $q_x$ on $m_1,m_2,\theta_{\rm b}$ and $\theta_{\rm r}$ (or $\mu,\sigma,\theta_{\rm b}$ and $\theta_{\rm r}$), and the numbers $N_{\rm{GW}}$ and $\bm{N_{\rm{EM}}}$, given by observations, according Equation~\eqref{eq10}, in principle. 
The constraints obtained from different realizations may be different. To consider this uncertainty, we generate $100$  realizations to get $\bm {N_{\rm{EM}}}$ and obtain the constraint. We find that the constraints obtained from $92$ realizations are within $1\sigma$ confidence intervals of those obtained from the one with the median value of $\bm{N_{\rm{EM}}}$ among the $100$ realizations. For demonstration, we only show the results for the one with the medium value of $\bm{N_{\rm{EM}}}$ in Table~\ref{mock} and Figures~\ref{fig:MCMC}-\ref{fig:par-dis} below. 

In this section, we show the reconstruction results and the ability of the counting method to limit the anisotropy parameters of kilonova for four cases.

First, for the ``Rubin'' scenario, we consider the observations of kilonovae by ignoring the effects of afterglows on the EM detection. Our calculations show that the input parameters in this case can be well reconstructed from the observations by the MCMC method, as seen in  Table~\ref{mock} (the 3rd row) and Figure~\ref{fig:par-dis} (the light blue lines). Apparently, the counting method can give a tight constraint on $\theta_{\rm b}$ but a poorer constraint on $\theta_{\rm r}$, if the distribution of $m_{\rm ej}$ is priorly known. For example, for the mock samples produced by assuming $m_{\rm ej}$ is logarithmic uniformly distributed in the range from $\log m_1$ and $\log m_2$, the constraints are $\log{m_1/M_\odot}=-3.98^{+0.65}_{-0.67}$, $\log{m_2/M_\odot}=-1.21^{+0.31}_{-0.25}$, $\theta_{\rm b}=0.60^{+0.23}_{-0.14}$\,rad, and $\theta_{\rm r}=0.91^{+0.45}_{-0.50}$\,rad, which are consistent with the input values for these parameters. If the prior on $m_{\rm ej}$ is not correct, then we would obtain biased constraints on $\theta_{\rm b}$ and $\theta_{\rm r}$. For example, for the same mock samples, if we adopt a log-normal distribution for $m_{\rm ej}$ as the prior, then we have $\theta_{\rm b}=0.50^{+0.17}_{-0.09}$\,rad and $\theta_{\rm r}=0.80^{+0.50}_{-0.47}$\,rad, which suggests the anisotropy radiation of kilonovae can still be reconstructed. However, the constraint on $\theta_{\rm b}$ is slightly biased from its input value though still tight.  

\begin{table*}
\centering
\caption{The expected number of the detected EM counterparts in different bands of mock BNS mergers in a realization, the input and the reconstructed values of the model parameters ($\log m_1$, $\log m_2$, $\theta_{\rm b}$, and $\theta_{\rm r}$) for four different cases. See detailed description of these different cases at the end of Section~\ref{sec:method}. }
\renewcommand{\arraystretch}{1.2} 
%
%
\makebox[\textwidth][c]{
\begin{tabular}{ccccccccccccccccc} \hline \hline
\multirow{2}{*}{Model} & \multicolumn{6}{c}{band}  & &  \multicolumn{4}{c}{input model parameter}  &  & \multicolumn{4}{c}{constraints}  \\ \cline{2-7} \cline{9-12} \cline{14-17}
 & u   & g    & r   & i   & z  & y & &$\log{m_1}$ &$\log{m_2}$ &$\theta_{\rm b}$ &$\theta_{\rm r}$ & &$\log{m_1}$ &$\log{m_2}$ &$\theta_{\rm b}$ &$\theta_{\rm r}$ \\ \hline
 Rubin      &0  &4  &5 &4  &2  &1   & &$-4$ &$-1$ &$\pi/6$ &$\pi/3$ &&$-3.98^{+0.65}_{-0.67}$ & $-1.21^{+0.31}_{-0.25}$& $0.60^{+0.23}_{-0.14}$& $0.91^{+0.45}_{-0.50}$\\
 Rubin+   &1 &16  &19 &16  &9 &3  & &$-4$ &$-1$ & $\pi/6$& $\pi/3$& &$-4.02^{+0.65}_{-0.66}$&
          $-1.14^{+0.29}_{-0.26}$&         $0.60^{+0.18}_{-0.12}$& $1.01^{+0.40}_{-0.48}$\\
 KN+AG\&info   &2 &17  &19 &16  &10 &3 &  &$-4$ &$-1$& $\pi/6$& $\pi/3$&&$-3.99^{+0.67}_{-0.66}$& $-1.16^{+0.29}_{-0.25}$&$0.61^{+0.19}_{-0.12}$& $0.97^{+0.42}_{-0.49}$\\          
KN+AG\&noinfo     &2 &17  &19 &16  &10 &3 &  &$-4$ &$-1$& $\pi/6$& $\pi/3$&&$-4.07^{+0.67}_{-0.60}$& $-1.18^{+0.30}_{-0.25}$ &$0.64^{+0.22}_{-0.15}$& $0.97^{+0.42}_{-0.46}$\\
\hline \hline
\end{tabular}
}
\begin{tablenotes}
\footnotesize
\item Notes: The first column (from top to bottom) lists the four different cases. The 2nd, 3rd, 4th, 5th, 6th, and 7th columns show the numbers of the detected EM counterparts in the u, g, r, i, z, and y bands estimated from mock realizations. For simplicity, the total number of BNS mergers detected by GW detectors is set as $N_{\rm{GW}}=3000$, which is compatible with the expected the detection rate of BNS mergers by the current and future GW detectors. The numbers in the rows denoted by ``Rubin'' and ``Rubin+'' are estimated by assuming that the EM counterparts are kilonovae only, and those in the rows denoted by ``KN+AG\&info'' and ``KN+AG\&noinfo'' are estimated by also considering the afterglow radiation in the model. The 8th, 9th, 10th, and 11th columns list the input values of the four model parameters, $\log m_1$, $\log m_2$, $\theta_{\rm b}$, and $\theta_{\rm r}$, for generating the kilonova light curves for mock BNS mergers. The last four columns list the reconstructed values of the above four parameters by using the counting method via the MCMC fitting. 
\end{tablenotes}
%
%
%
\label{mock}
\end{table*}

\begin{figure*}
\centering
\includegraphics[width=0.5\textwidth]{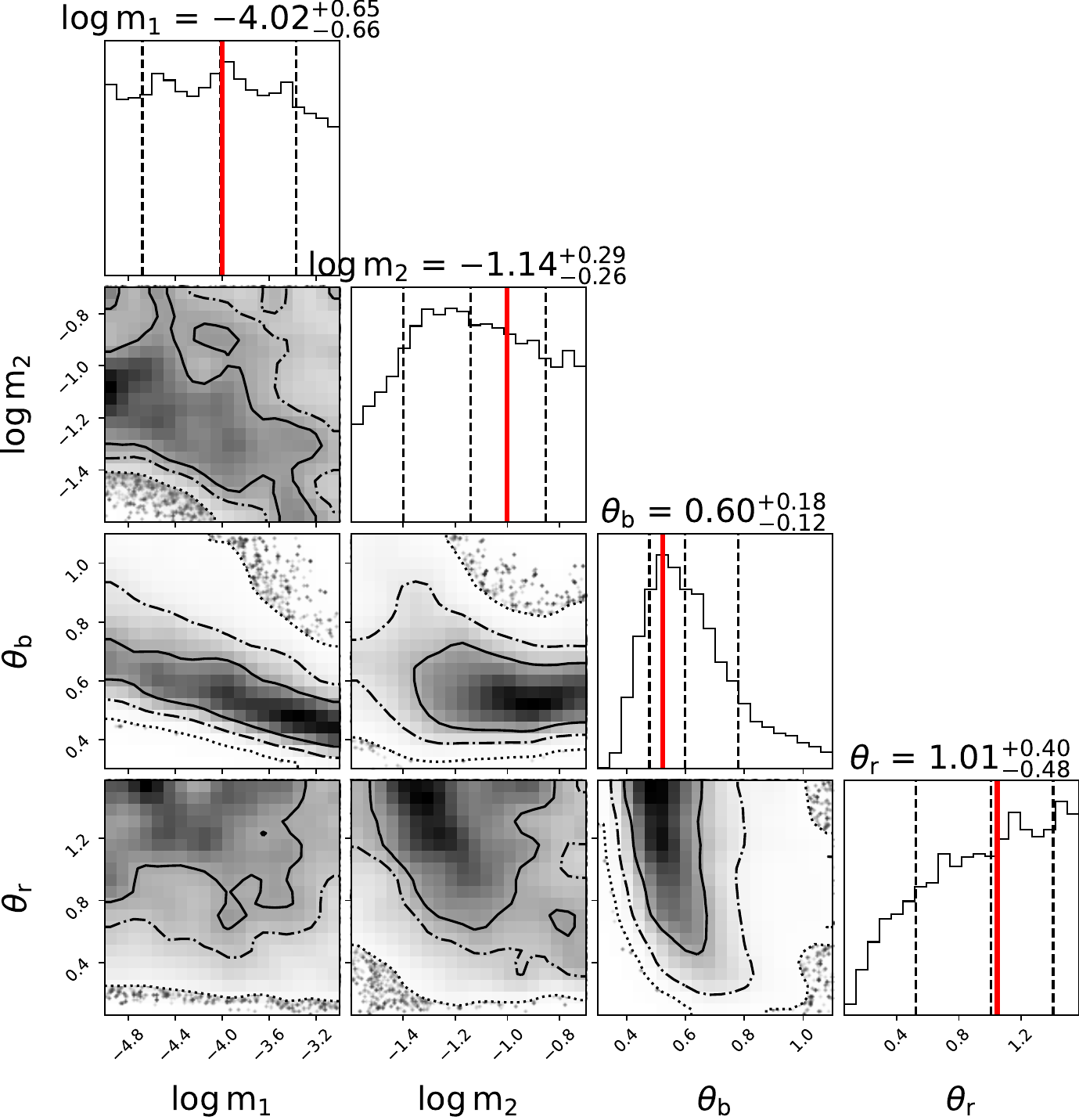}
\caption{
The probability distributions of the reconstructed model parameters obtained by the MCMC method for the ``Rubin+'' case. The input values of the model parameters used to generate mock observations are $\log (m_1/M_\odot)=-4$, $\log (m_2/M_\odot)=-1$, $\theta_{\rm b}=\pi/6$, and $\theta_{\rm r}=\pi/3$, as indicated by the red vertical line in each top panel from left to right. The solid, dash-dotted, dotted lines show the $50\%$, $80\%$, $98\%$ credible intervals respectively. 
}
\label{fig:MCMC}
\end{figure*}

\begin{figure}
\centering
\includegraphics[width=0.48\textwidth]{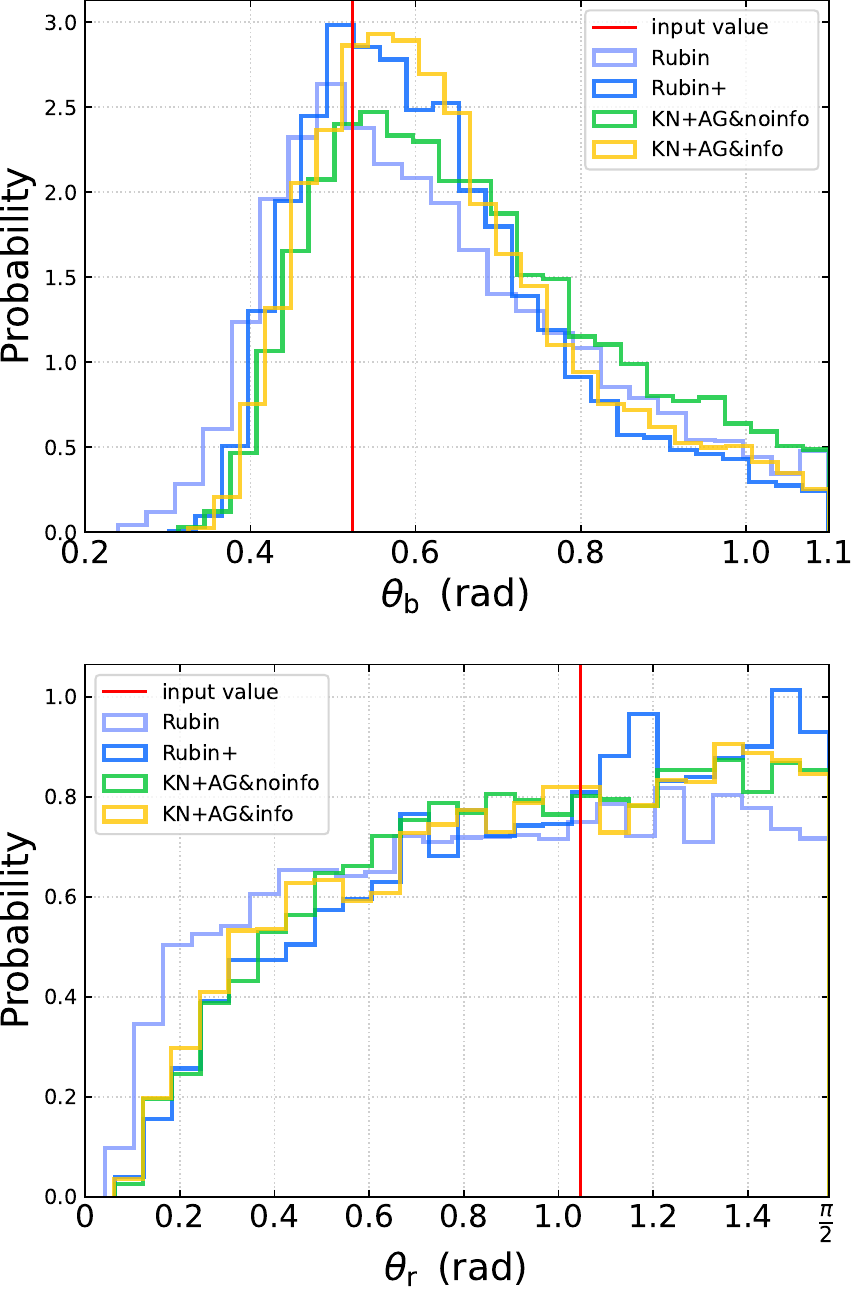}
\caption{
The Probability distributions of $\theta_{\rm b}$ and $\theta_{\rm r}$ for five different situations which are described in Section~\ref{sec:method}. Different color histograms represent different situations. The vertical red lines are the input values to generate mock observations. 
}
\label{fig:par-dis}
\end{figure}

It is expected that a survey deeper than Rubin would detected more kilonovae, and thus give better constraints on the model parameters by the counting method. To see such an effect, we consider the  ``Rubin+'' case, which is assumed to be similar to Rubin but with a limiting magnitude one magnitude deeper than that of Rubin. The results of the constraints obtained from the mock samples generated for this case are listed in Table~\ref{mock} (the 4th row) and shown in Figure~\ref{fig:MCMC}, and the one dimensional probability distributions for the reconstructed model parameters $\theta_{\rm b}$ and $\theta_{\rm r}$ are also shown in Figures~\ref{fig:MCMC} and \ref{fig:par-dis}. As seen from Figures~\ref{fig:MCMC},  $\theta_{\rm b}$ and $\theta_{\rm r}$ can be well reconstructed, and the latter is less well reconstructed comparing with the former, if we adopt a prior distribution of $m_{\rm ej}$ as uniformly distributed in the logarithmic scale similar to the input ones. As expected, ``Rubin+'' can give stronger constraints on the kilonova anisotropy radiation, i.e., a well reconstructed $\theta_{\rm b}$ with a smaller scatter comparing with that by ``Rubin''. If assuming a log-normal distribution for $m_{\rm ej}$, then we still obtain a slightly biased constraint on $\theta_{\rm b}$ as $\theta_{\rm b}=0.49^{+0.10}_{-0.07}$\,rad, and this result can still give a tight constraint on it.

The observational light curves of a BNS merger may be the combination of both the kilonova and the afterglow radiation, especially when the viewing angle of the BNS merger is close to face on. In the case of a small viewing angle (or close to face on), the effect of afterglow is more significant. Therefore, one may need to consider such afterglow radiation when using the counting method to constrain the anisotropy of kilonova radiation. Therefore, we also calculate the light curves of the afterglow multi-band radiation for each BNS merger. We consider two different cases by invoking the effect of afterglow on the detection of BNS EM counterparts. The first case is for the ``KN+AG\&info'' case listed in Section~\ref{sec:method}, in which we adopt the KN+AG model to constrain the kilonova model parameters with the counting method by assuming the afterglow signatures are identified, and the results are listed in the 5th row as ``KN+AG\&info'' in Table~\ref{mock} and shown as yellow lines in Figure~\ref{fig:par-dis}. The model parameters $\theta_{\rm b}$ and $\theta_{\rm r}$ for kilonovae can still be well reconstructed, i.e., $\theta_{\rm b}=0.61^{+0.19}_{-0.12}$\,rad, $\theta_{\rm r}=0.97^{+0.42}_{-0.49}$\,rad, which means that the anisotropy of the kilonovae can be revealed via the counting method. The second case is for the ``KN+AG\&noinfo'' case listed in Section~\ref{sec:method}, in which we only adopt the KN model to constrain the kilonova model parameters with the counting method by ignoring the afterglow signals, though the afterglow radiations are included in the mock observations. For this situation, the obtained constraints on $\theta_{\rm b}$ and $\theta_{\rm r}$ for the kilonova model are slightly larger than other cases, as seen from the 6th row in Table~\ref{mock} and the green lines in Figure~\ref{fig:par-dis}. To obtain robust constraint on the aniostropy of kilonova radiation and/or the kilonova model, one needs to carefully consider the effect of the afterglow radiation on the light curves of EM counterparts though it is small.

\begin{figure}
\centering
\includegraphics[width=0.48\textwidth]{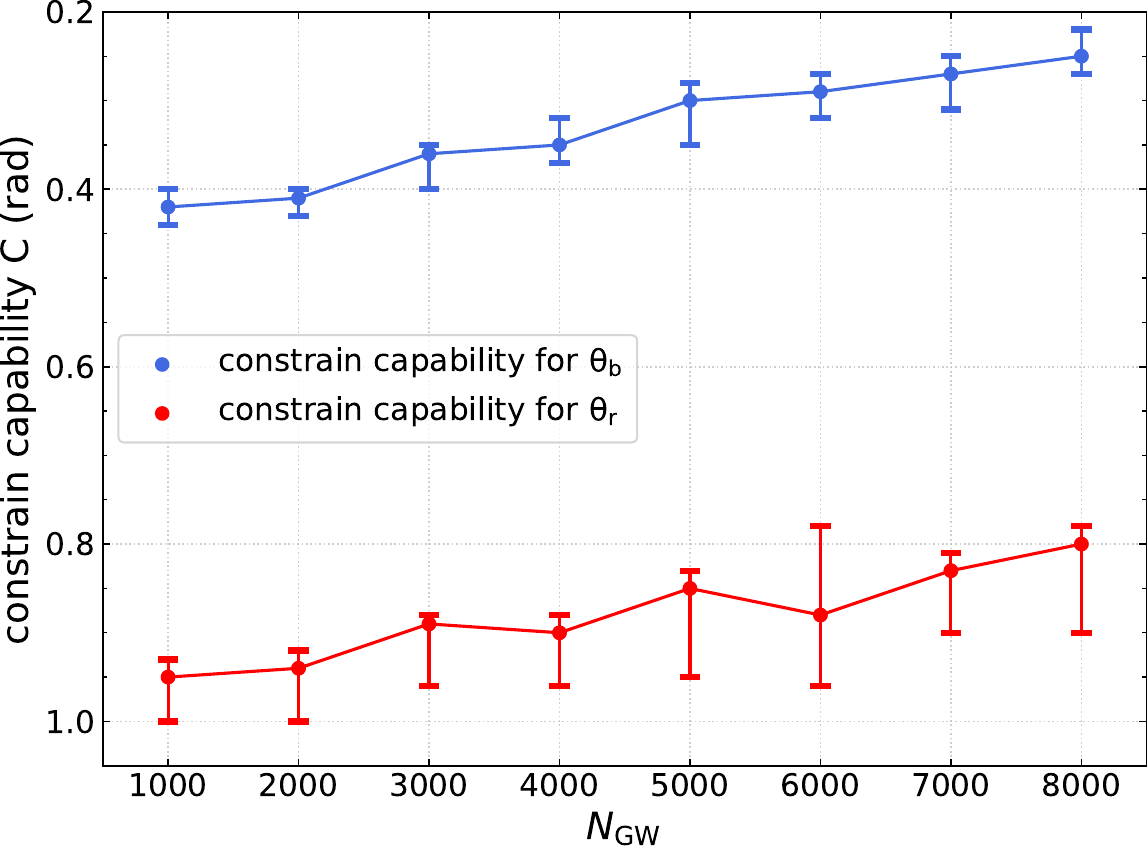}
\caption{
The constraining capabilities $C$ of the counting method for $\theta_{\rm b}$ and $\theta_{\rm r}$ as a function of the detected number of BNS merger GW events. Here $C$ is the width of the probability distribution (from the 84\% to the 16\% quantiles) for the reconstructed parameter. Using the same input parameters shown in Table~\ref{mock} and consider the ``Rubin+'' case, we carry out $30$ realizations of mock observations for each given $N_{\rm{GW}}$ to obtain the corresponding $\bm{N_{\rm{EM}}}$. The solid lines and filled circles represent the mean values of $C$ and its errorbars obtained by the reconstruction for all the realizations for these events.
}
\label{fig:C_NGW}
\end{figure}

In the above calculations, we set $N_{\rm{GW}}= 3000$ as an example. The expected detection rate of BNS mergers is $\sim 10^1-10^3$ per year in O5 \citep{Petrov2022, Weizmann2023} and $\sim 10^4-10^5$ by the third generation ground-based GW detectors (ET and CE) \citep{Belgacem2019, 2022Singh, mahao2022}. The detection number of BNS mergers by GW detectors increases with time. Here we further calculate the possible constraints that can be obtained by using different numbers of detected BNS mergers by GW detectors. Figure~\ref{fig:C_NGW} illustrates the constraining capability of the counting method as a function of the number of detected BNS merger GW events. Here the constraining capability is defined as $C$, which refers to the width of the probability distribution of the constrained parameter, specifically the width from the $84\%$ to $16\%$ of the distribution. The stronger the constraining capability achieved through the counting method, the narrower the parameter distribution becomes, i.e., a smaller value of $C$. Therefore, the definition of $C$ can in principle reflects the constraining capability. As expected, the constraints would be improved with increasing number of GW detected BNS mergers (correspondingly increasing number of detected EM counterparts).

\section{Conclusions and Discussions}
\label{sec:con}

In this work, we have investigated the feasibility of using the counting method to statistically constrain the kilonova model, especially the anisotropy radiation. The required information in this method are mainly the detection number of GW events and the numbers of their EM counterparts detected in different bands by a survey with given limiting magnitudes and searching strategy. By generating mock samples with a simple two-component kilonova model, performing mock observations, and doing MCMC fittings to reconstruct the model parameters, we demonstrate that the kilonova model can be well constrained and the anisotropy of kilonova radiation, especially represented by the opening angle of the blue components, can be revealed by future observations via the counting method. The advantage of this counting method is that it needs only a little information and can give global constraints on the model parameters for the whole population by using both the detection and non-detection of kilonovae produced from the BNS mergers observed by GW detectors. 

The constraints on the opening angles of both blue and red components obtained from the counting method may depend on the prior about the ejecta mass distribution of BNS mergers. If this prior deviates from the true one, the obtained constraints may be biased away from the true one. In principle, if there are a fraction of the detected kilonovae have multi-band multi-epoch detailed observations, one may be able to extract information about the ejecta mass distribution and then use it as the prior for the counting method. We note that \citet{Farah2020} and \citet{Biscoveanu2020} have investigated about the prior information that can be obtained from the two observed BNS merger events, GW170817 and GW190425. We expect that one could have some prior information about the BNS merger events via both the GW and EM signals, and thus enables tight constraints on the anisotropy of kilonova radiation from the counting method.

In our calculations, we fix $\chi$, $\kappa$ and $v_{\rm ej}$, and assume that these parameters are consistent with the results given by the fitting results of \citet{Villar2017} for GW170817, the only BNS merger that have both GW and EM observations. In principle, these parameters can have wide distributions among different kilonovae. Ignoring the wide-range distributions of these parameters may lead to bias in the constraining of the anisotropy of kilonova radiation and the kilonova model parameters. However, if we have prior information about the distributions of these parameters, which can be included in the kilonova modelling, then we would still obtain tight constraints on the anisotropy parameters (including $\theta_{\rm b}$ and $\theta_{\rm r}$). It is expected that there will be a number of kilonovae and afterglows associated with the BNS merger GW events being detected with detailed LCs, similar as GW170817, in the near future. 
Specifically, in the O4 and O5 operation runs of LIGO, Virgo, and KAGRA \citep{Abbott2020}, and the era of the third-generation GW detectors, the detection rates of kilonovae are expected to be $\sim1-5$, $\sim10-50$, and $\sim1000$ per year, respectively \citep[e.g.,][]{Petrov2022, Weizmann2023, Shah2023, Colombo2022, Zhu2021b}. A significant fraction of these events may have multi-band light curves as those obtained for GW170817. 
With these detections, the detailed EM observations should enable precise extraction of the kilonova parameters like $\chi$, $\kappa$, and $v_{\rm ej}$, and thus give their distributions. One could take these distributions into the kilonova model when using the counting method to constrain the anisotropy of kilonova radiation and the kilonova model, and thus avoid the possible bias in those constraints obtained by fixing $\chi$, $\kappa$, and $v_{\rm ej}$, at the values as that obtained from GW170817. For simplicity, we do not incline to consider this complication but defer it to future works. 

It should be emphasized that for a given set of model parameters, the lower BNS detection rate may result in fluctuations in the number of $\bm{N_{\rm{EM}}}$ across different bands due to small number statistics.
However, with the high detection rate  $\sim 10^4-10^5$ of BNS mergers per year by the third generation ground-based GW detectors (ET and CE) \citep{Belgacem2019, 2022Singh, mahao2022}, one would expected that the small number statistics would be overcome.

Furthermore, the kilonova model adopted in this paper is a simplified two-component model. A kilonova model consists of three components may provide a better fit to the kilonova light curves of AT2017gfo \citep[][]{Cowperthwaite2017}. In addition, many processes that could contribute  additional energy to kilonova, such as the fallback accretion \citep[][]{Rosswog2007, Rossi2009, ZCY2023} or the magnetic winds from a long-lived magnetar remnant \citep[][]{Yu2013, Gao2015, Radice2018, MetzgerandPiro2014}, are not considered here. These all deserve further study. Nevertheless, it is expected that the counting method can still provide a global constraint on the kilonova model and the anisotropy of the kilonova radiation, even with a more complicated kilonova model.
  
\section*{Acknowledgements}
We thank the referee for insightful comments and helpful suggestions. SQZ thanks Dr. Weiyang Wang for helpful suggestions and comments. This work is partly supported  by the National Natural Science Foundation of China (Grant No. 12273050, 11991052, 11690024), the Strategic Priority Research Program of the Chinese Academy of Sciences (grant no. XDB0550300), and the National Key Program for Science and Technology Research and Development (Grant No. 2020YFC2201400).

\section*{Data Availability}
The data underlying this article will be shared on reasonable request to the corresponding author.



\bibliographystyle{mnras}
\bibliography{ref} 

\bsp	
\label{lastpage}
\end{document}